# On the environment-destructive probabilistic trends: a perceptual and behavioral study on video game players

Quan-Hoang Vuong [1,2,3,*], Manh-Toan Ho [1,3,*], Minh-Hoang Nguyen [1,3], Thanh-Hang Pham [4], Hoang-Anh Ho [3], Thu-Trang Vuong [5], Viet-Phuong La [1,3]

[1] AISDL, Vuong & Associates, Dong Da, Hanoi 100000, Vietnam

[2] Centre Emile Bernheim, Université Libre de Bruxelles, 1050 Brussels, Belgium

[3] ISR, Phenikaa University, Ha Dong, Hanoi 100803, Vietnam

[4] Faculty of Management and Tourism, Hanoi University, Km9, Nguyen Trai Road, Thanh Xuan, Hanoi 100803, Vietnam

[5] Sciences Po Paris, 27 Rue Saint-Guillaume, 75007 Paris, France

* Correspondence: Quan-Hoang Vuong (qvuong@ulb.ac.be); Manh-Toan Ho (toan.homanh@phenikaa-uni.edu.vn)

## Abstract

Currently, gaming is the world's favorite form of entertainment. Various studies have shown how games impact players' perceptions and behaviors, prompting opportunities for purposes beyond entertainment. This study uses Animal Crossing: New Horizons (ACNH)—a real-time life-simulation game—as a unique case study of how video games can affect humans' environmental perceptions. A dataset of 584 observations from a survey of ACNH players and the Hamiltonian MCMC technique has enabled us to explore the relationship between in-game behaviors and perceptions. The findings indicate a probabilistic trend towards exploiting the in-game environment despite players' perceptions, suggesting that the simplification of commercial game design may overlook opportunities to engage players in pro-environmental activities.

*\*\*\**

*THIS IS A NON-PEER REVIEWED MANUSCRIPT*

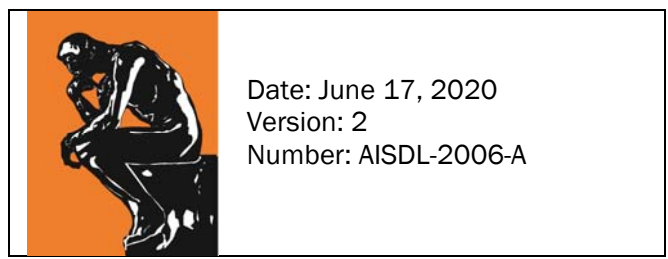

Date: June 17, 2020
Version: 2
Number: AISDL-2006-A

# On the environment-destructive probabilistic trends: a perceptual and behavioral study on video game players

## Introduction

Video games are suggested to affect players' perceptions and behaviors (Bejjanki et al., 2014; Franceschini et al., 2017; Hwang & Lu, 2018). For instance, action video games have been reported to improve dyslexia children's reading abilities (Franceschini et al., 2017) or support perceptual learning ability (Bejjanki et al., 2014). Therefore, researchers have shown interest in using such games to educate the public about environmental issues such as climate change or global warming (Kwok, 2019). Early results have indicated that such well-intentioned games, like *World Climate* simulation, can significantly help players gain knowledge about causes and effects of climate change, its urgency and motivate them to be more proactive (Rooney-Varga et al., 2018). However, these "serious games" designed for purposes beyond entertainment are much less popular than commercial games (DeWeerdt, 2018).

Totally, the game industry has over 2.7 billion players and generates almost 160 billion dollars in revenue (Wijman, 2020). Given this huge market size, they have a strong potential to become a viable tool that serves multi-purposes, including shaping players' environmental perceptions (McCartney, 2013; Trista & Sam, 2019). As such, commercial games may also have similar impacts even more naturally than the somewhat coercive gamification of educational content.

As a typical example, Nintendo's Animal Crossing: New Horizons (ACNH), a life-simulation game, has attracted around 11 million players since its release in March (*Guardian* 13-May-2020: https://bit.ly/2zBCl4Z). Its distinctive freedom and environmental interaction have offered an unprecedented experience. Players can build and personalize their own islands by customizing their ecosystem and community. They frequently engage in environmental activities like planting trees and flowers, gardening, fishing, or hunting for bugs and fossils. The interaction is so lifelike that People for the Ethical Treatment of Animals (PETA) has issued a guideline for proper behaviors towards the in-game animals (*PETA* 2020: https://bit.ly/3exh3nR).

With a dataset of 584 responses from a comprehensive survey of ACNH players about their environmental perceptions and behaviors, this study explores the following questions:

1. Do people with different socio-demographic characteristics have different perceptions of humans' rights to modify the natural environment?
2. Do people with different socio-demographic characteristics have different in-game behaviors?
3. What is the correlation between in-game behaviors and environmental perceptions of the players?

## Materials and Methods

### Samples

On 15 May 2020, we started posting questionnaires on Facebook groups, Reddit, and Discord servers with informed consent collected at the beginning. The participants answered questions about four aspects: 1) Socio-demographic factors, 2) Game playing experience, 3) In-game behaviors, 4) Environmental perceptions. As of 28 May 2020, 584 responses from 26 countries and regions with high ethnic diversity were collected and used for analysis (Ho, Nguyen, Pham, La, & Vuong, 2020).

The dataset contains variables regarding players' information (sex, educational level, geographical location), environmental perceptions (whether the participant agrees with humans having the rights to modify natural environment to suit their needs), and players' in-game behaviors (whether players are likely to collect animals/plants).

## Methodology

A Bayesian analysis—the Hamiltonian Markov chain Monte Carlo (MCMC) technique—is performed to investigate the probabilistic trends of video game players' environmental perceptions and in-game behaviors based on their gender, educational level, and geographical location. The correlation between the environmental perceptions and in-game behaviors of players is explored using Bayesian multilevel modeling. The **bayesvl** and **rstan** R packages are employed concurrently for performing all the analyses (Q.-H. Vuong et al., 2020), and the **loo** package is used for estimating weights of regression models. Technical treatments are provided in Supplementary Files at: https://osf.io/p8u9c/

## Robustness Check

In this article, we use a 2-pronged approach to robustness check: tweaking the priors of each model; and performing sample-size sensitivity check with two datasets with 503 and 584 observations, respectively. Both approaches to robustness check show a low sensitivity to change in priors or sample size (See Table 1a).

**Table 1.** Robustness check

| a) Priors sensitivity and sample-size sensitivity | | | | | | |
|---|---|---|---|---|---|---|
| Dataset | Environmental perception | | | In-game behaviors | | |
| | Parameter | mean | sd | Parameter | mean | sd |
| | *Prior: Normal(0.3, 10)* | | | | | |
| Dataset [584] | Theta_C2 | 0.47 | 0.03 | theta_animal | 0.97 | 0.01 |
| | | | | theta_Tree | 0.96 | 0.01 |
| Dataset [503] | Theta_C2 | 0.48 | 0.03 | theta_animal | 0.97 | 0.01 |
| | | | | theta_Tree | 0.96 | 0.01 |
| | *Prior: Normal(0.6, 10)* | | | | | |
| Dataset [584] | Theta_C2 | 0.47 | 0.02 | theta_animal | 0.97 | 0.01 |
| | | | | theta_Tree | | |
| Dataset [503] | Theta_C2 | 0.48 | 0.03 | theta_animal | 0.97 | 0.01 |
| | | | | theta_Tree | 0.96 | 0.01 |
| | | | | | | |
| b) Weight comparison between model [1] and [2] | | | | | | |
| Regression Model | WAIC | Bayesian stacking | Pseudo-BMA with Bayesian bootstrap | Pseudo-BMA without Bayesian bootstrap | | |
| Model [1] | 0.417 | 0.574 | 0.497 | 0.510 | | |
| Model [2] | 0.583 | 0.426 | 0.503 | 0.490 | | |

# Results

Firstly, the associations between gender, educational level, and geographical location of ACNH players and their environmental perceptions are explored. In Figure 1a, male players are more likely to agree with exploiting the environment for human benefit ($\theta_{Male}$=0.67, SD=0.04), while the probabilistic trend of female players presents a contradicted result ($\theta_{Female}$=0.36, SD=0.03). Environmental perceptions also vary across geographical locations and educational levels. Geographically, players from US/Canada ($\theta_{US/Canada}$ = 0.53, SD=0.03) have a contradicted viewpoint with those from Europe ($\theta_{EU}$=0.28, SD=0.06). Regarding educational background, undergraduate students have a higher propensity to agree with the statement ($\theta_{Undergraduate}$=0.49, SD=0.05), while secondary students are still forming their environmental perceptions ($\theta_{Secondary}$=0.44, SD=0.13).

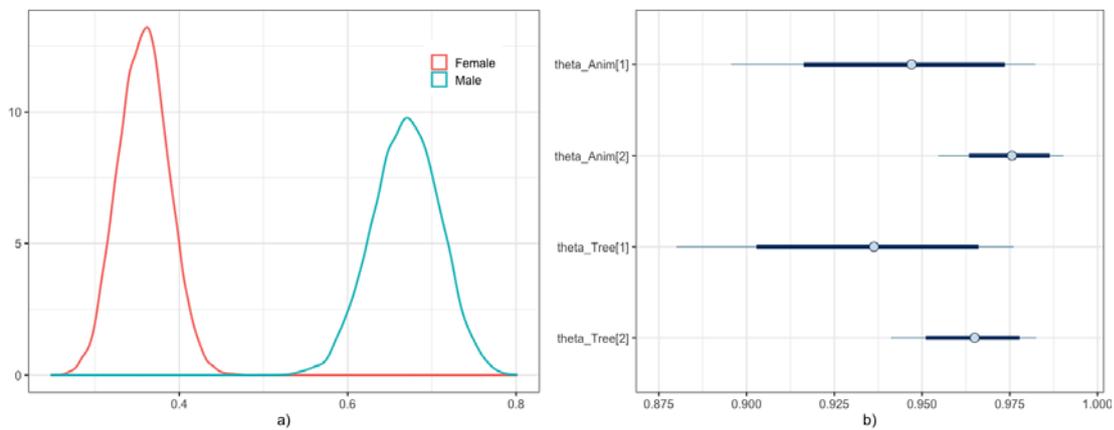

**Figure 1:** a) Probability distribution of environmental perception by sex; and b) Probability distribution of behaviors towards animals and plants by sex.

For the second research question, unlike their perceptions, ACNH players' in-game behaviors show a trend towards exploiting the environment, regardless of their gender, educational level, or geographical location. Specifically, even though female and male players have contradicted perceptions in the previous model, Figure 1b suggests that their in-game behaviors have an equally significantly high level of collecting animals ($\theta_{Animal\_Male}$=0.95, SD=0.02; $\theta_{Animal\_Female}$=0.98, SD=0.01) or plants ($\theta_{Tree\_Male}$=0.94, SD=0.03; $\theta_{Tree\_Female}$=0.97, SD=0.01).

Finally, we created two combined models [1] and [2] to investigate the correlation between environmental perceptions and in-game behaviors. The results continue to affirm a juxtaposition between video games players' environmental perceptions and their in-game behaviors. In model [1], which considers catching animals and collecting plants separately (Fig.2a), players who catch animals ($\beta_{Animal}$=2.22, SD=1.39) have a higher correlation with accepting nature-exploiting mindset. On the contrary, collecting plants show negligible impacts on players' perceptions ($\beta_{Tree}$=0.03, SD=0.77). However, the model [2], which considers the interaction of both activities (Fig.2b), reveals that players are going to exploit the environment even more when participating in both ($\beta_{Animal\_Tree}$=5.75, SD=4.44). The weight comparison in Table 1b of the two models also suggests that the model with interaction (WAIC=0.583) is more probable than the one without interaction (WAIC=0.417).

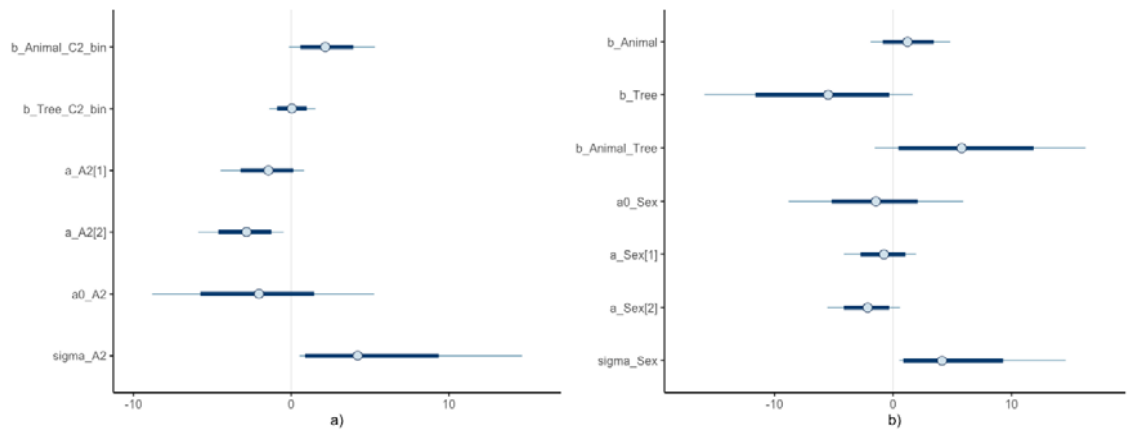

**Figure 2:** a) Model 1's probability distribution of parameters; b) model 2's probability distribution of parameters.

## Discussion

The limitation of in-game choices is somehow similar to the constraints that a person might face in exercising pro-environmental behaviors. For instance, farmers might love nature, but the reality of life can force them to exploit the environment to make ends meet. In ACNH, we suspect that players' behaviors may result from the simplification of game design. A core mission at the beginning of the game makes players pay a mortgage for their houses by catching animals or collecting fruits and flowers. This conditions players into exploiting the environment for their benefits. Video games should have done better, especially when commercial games have more sophisticated designs with technological advancements. To illustrate, the causes and effects of climate change can become an integral part of the ACNH ecological system by having consequences for excessive behaviors.

Currently, environmental sustainability is not a visible element in game design. However, through the interactivity in a virtual world, game designers can make meaningful impacts on players' perceptions about this issue, thus indirectly creating a sustainable future for themselves. In fact, the United Nations and other international organizations have joined forces with game companies such as Sony or Microsoft to produce greener physical products (United Nations, 2019). We also suggest developing a guideline to encourage "UN-compliance games" to incorporate pro-environmental elements in their core design. To design such gaming experiences, the industry needs more evidence from reality. This is what our study hopes to provide for the industry, as well as science and the society (Q. H. Vuong, 2018, 2020).

Our data collection process suggests that learning about game players is a hard but rewarding process. In the future, the increasing numbers of activities in the virtual world will produce a large stream of data for scientists to learn from. As the UN calls for a measurement of how in-game green elements impact players (Trista & Sam, 2019), the treatment of these data will be valuable to the game industry, international organizations, policymakers, and other stakeholders in making games beyond entertainment. Video games are no longer a subject of criticism for its influence on players (Kühn et al., 2019), but an opportunity to educate younger populations about the risks that the world is facing through a computer-generated mindsponge process (Vuong & Napier, 2015).

## Acknowledgments

The authors sincerely thank administrators and members of Facebook groups, subreddit, and Discord servers about Animal Crossing: New Horizons.